\documentclass[aip,jap,preprint]{revtex4-1}

\date{\today}

\usepackage{graphicx}
\usepackage{amssymb}
\usepackage{dsfont}
\usepackage{bm}

\begin{document}

\title[Transition from positive to negative
magnetoresistance in semiconductor nanowire]%
{Transition from positive to negative
magnetoresistance induced by a constriction
in semiconductor nanowire}

\author{M. Wo{\l}oszyn}
\author{B.J. Spisak}
\email[Electronic address: ]{bjs@agh.edu.pl}
\author{P. W\'ojcik}
\author{J. Adamowski}

\affiliation{AGH University of Science and Technology, Faculty of
Physics and Applied Computer Science, al.~Mickiewicza 30,
30-059~Krakow, Poland}

\begin {abstract}
We have studied the magnetotransport through an indium antimonide (InSb)
nanowire grown in [111] direction,
with a geometric constriction and in an external magnetic field applied 
along the nanowire axis.
We have found that the magnetoresistance is negative for the narrow constriction,
nearly zero for the constriction of some intermediate radius, and takes on 
positive values for the constriction with the radius approaching that of the 
nanowire.
For all magnitudes of the magnetic field, the radius of constriction at which 
the change of the magnetoresistance sign takes place has been found to be almost 
the same as long as other geometric parameters of the nanowire are fixed.
The sign reversing of the magnetoresistance is explained as a combined
effect of two factors: the influence of the constriction on the transverse
states and the spin Zeeman effect.
\end{abstract}


\maketitle

\section{\label{sec:intro}Introduction}

One of the common theoretical strategies used to investigate the electronic 
transport in the solid state systems, and in particular in the nanostructures,
is based on the calculations of the response to an external perturbation due to 
the electromagnetic field or temperature. 
Within the framework of the linear response theory, the reaction of the electron 
system is described by the relevant kinetic coefficients. 
For this reason, the studies of the magnetotransport properties of the 
nanowires and other nanostructures can be based on the analysis of the electron 
response to a suitably oriented magnetic field in terms of the 
magnetoresistance (MR), which is defined as the relative change of the 
resistance due to the applied magnetic field. 
The importance of this phenomenon results in
the numerous practical applications, which include hard disks, memories, and
various sensors.

It is worth recalling that the semiclassical theory of the galvanomagnetic 
phenomena predicts the positive MR with the $B^2$-dependence for the weak 
magnetic field $B$, and saturation of MR for the strong magnetic 
field.\cite{Dresden_RMP33p265y1961, Ziman_EandPh2001, Kaganow_PR372p445y2002}
A certain deviation from the semiclassical theory has been found
experimentally in a number of different systems. 
For example, the quasi-linear $B$-dependence of MR is observed in the limit of 
the high magnetic field in the bulk $n$-type InSb at liquid-nitrogen 
temperature,\cite{Frederikse_PR108p1136y1957, Hu_Nature7p697y2008} and
a similar dependence of MR on the magnetic field is observed in silver 
chalcogenides.\cite{Lee_PRL88p066602y2002}
The explanation of the non-saturating properties of MR can be based on the 
large spatial fluctuations in the conductivity of the narrow-gap semiconductors, 
due to the inhomogeneous distribution of silver 
ions.\cite{Hu_PRB75p214203y2007, Parish_Nature426p162y2008}
It has been also shown that the large positive MR is induced by the 
quasi-neutrality breaking of the space-charge effect in 
Si.\cite{Delmo_Nature457p1112y2009}
The large positive MR has been also reported by Schoonus et al. in Boron-doped 
Si--SiO$_2$--Al structures.\cite{Schoonus_PRL100p127202y2008}

Some of the available experimental data show that MR can be negative. 
In the disordered systems it can be explained by the weak 
localization theory, which predicts the negative MR with 
the $\sqrt{B}$-dependence.\cite{Kawabata_JPSJ49p628y1980,
Altshuler_JETP54p411y1981,Morgan_JN-CS270p269y2000}
Moreover, in the disordered systems, the sign of MR can be affected by the 
spin-orbit interaction.\cite{Hikami_PTP63p707y1980,Fukuyama_JPSP50p2131y1981, 
Chakravarty_PR140p193y1986,Spisak_pssb242p1460y2005}
Nevertheless, in organic semiconductor devices the transition between 
positive and negative MR due to the applied voltage and temperature has been 
observed,\cite{Bloom_PRL99p257201y2007, Gu_APL102p212403y2013}
but the microscopic origin of this effect is still unclear. 
In Ref.~\onlinecite{Hu_Nature6p985y2007}, it has been shown that 
MR can be changed from positive to negative by adjusting 
the dissociation and charge reaction in excited states by changing the 
bipolar charge injection in the organic LED. 
A similar change of MR sign is possible in the bilayer graphene, where
the gate voltage induces switching from 
the negative to the positive MR.\cite{Chen_PB406p785y2011}
The mechanism responsible for the switching is related to the strong 
contribution from the magnetic-field modulated density of states together 
with the weak localization effects.
Such mechanisms are also responsible for the transition from the positive to the
negative MR in the double-walled carbon 
nanotubes,\cite{Fedorov_PRL94p066801y2005} although Roche and Saito demonstrate 
that MR in such carbon nanotubes can be either positive or negative, depending 
on the chemical potential and the orientation of the magnetic field with respect 
to the nanotube axis.\cite{Roche_PRL87p246803y2001} 
All these examples prove that predicting the sign of MR and its
field--dependence in the nanostructures is a nontrivial task.
 
The nanowires made of InSb are very interesting 
nanosystems for investigations of modern concepts in nanoelectronics, and 
spintronics in particular. 
For example, 
in the presence of the magnetic field, the phase coherent transport is observed
in InSb nanowires at low temperatures.\cite{Yao_APL101p082103y2012}
Besides,
the quantization of the conductance in the nanosystems has been experimentally 
confirmed more recently,\cite{Weperen_NL13p387p2013} although this quantum
effect in the $3D$ nanowires has been predicted much 
earlier.\cite{Scherbakov_PRB53p4054y1996, Vargiamidis_JAP92p302y2002, 
Waalkens_PRB71p035335y2005}
The quantization of the conductance is difficult to observe in the real 
nanowires due to the presence of structural and substitutional disorder, 
and because of the boundary roughness.\cite{Weperen_NL13p387p2013}
This stems from the fact that scattering of conduction electrons on impurities 
or on structural imperfections results in the change of momentum 
(the momentum relaxation), which leads to smearing of the step-like form of 
the electric conductance.

In this paper, we study the influence of the spin degree of freedom on 
the magnetotransport properties of the three-dimensional InSb 
nanowire with a constriction placed at the half-length of the nanowire, and
in the presence of the magnetic field directed along the axis of the nanowire.
Utilization of the MR effect in the nanowires, which can possibly replace
devices of larger extents, may be seen as an opportunity to enable high 
sensitivity, while the small power consumption is ensured.
Recent studies on nonmagnetic III-V nanowires suggest such possibility
for future high-density magneto-electric devices, compatible with 
commercial silicon technology.\cite{Li2015}
The available experimental reports show that the change of MR sign can be
related to the applied gate voltage in a number of different materials,
including organic semiconductors\cite{Bloom_PRL99p257201y2007} and
carbon nanotubes\cite{Fedorov_PRL94p066801y2005}.
Electric control of MR, both its sign and magnitude, was also reported 
in the case of InP nanowires.\cite{Zwanenburg2009}
Our calculations show that this effect can be also induced by the presence
of the constriction in the nanowire.

The paper is organized as follows. 
In Sec.~\ref{sec:theory}, we present the three-dimensional model of the 
semiconductor nanowire with the geometric constriction, and 
introduce the theoretical 
method used to investigate the magnetotransport properties of this 
nanostructure in the coherent regime of the electronic transport. 
Sec.~\ref{sec:results} contains the results of calculations and their 
discussion, and Sec.~\ref{sec:concl} -- the conclusions.

\section{\label{sec:theory}Theory}

\begin{figure}
\begin{center}
\includegraphics[width=0.50\textwidth]{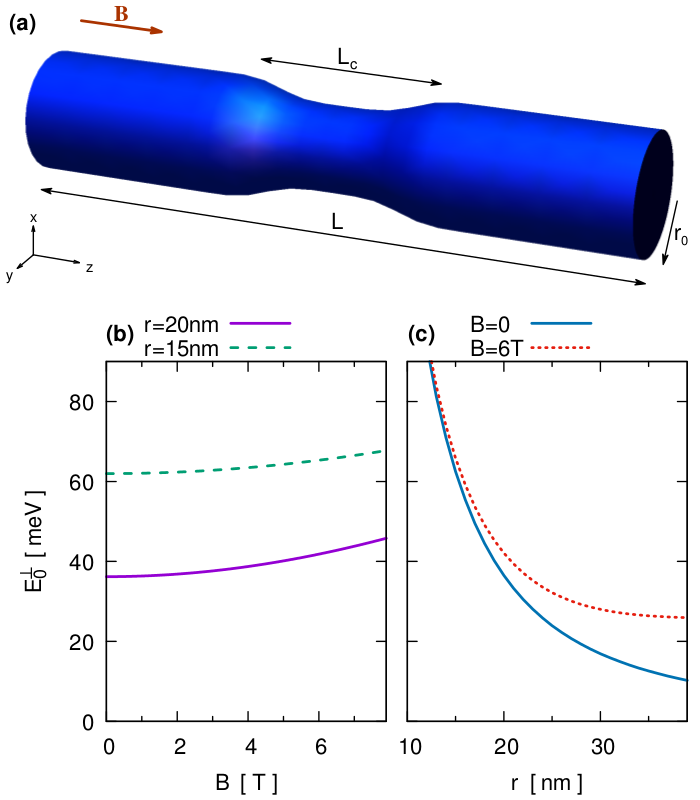}
\caption{
(a)~Schematic of the nanowire with a single constriction.
$L$ is the length of the nanowire, the region of constriction
has length $L_c$, and $r_0$ is the radius of the nanowire outside the 
constriction.
(b)~Dependence of the transverse eigenenergy $E_0^{\perp}$ on the magnetic
field at two distinct $z$-coordinates, at which the radius of the nanowire
is equal to 20~nm (solid lines) or 15~nm (dashed lines); $r$ is the radius
of the nanowire at $z$.
(c)~Same as (b), but as a function of the radius of the cross-section,
at $B=0$ and $B=6$~T.
}
\label{fig:model}
\end{center}
\end{figure}

We consider the InSb nanowire grown in [111] direction, and with a constriction
in the middle of its length, as
presented schematically in Fig.~\ref{fig:model}(a).
The nanowire is modeled as a cylindrical rod, which has a negligible effect
on the electronic transport because
we concentrate on the geometric and material parameters, for which only 
the ground transverse state plays a role.\cite{Woloszyn_JP-CM26p325301y2014}

Within the effective mass approximation, the $2\times 2$ conduction band 
Hamiltonian has the form
\begin{equation}
\hat{\mathcal{H}} =
\bigg[ \frac{\hat{\bm\pi}^2}{2m^{\ast}} + U_{conf}(\mathbf{r}) + eFz \bigg]
\hat{\mathds{1}}
+ \hat{\mathcal{H}_Z} + \hat{\mathcal{H}_D} + \hat{\mathcal{H}_R}.
\label{Hamiltonian}
\end{equation}
The kinetic momentum is defined by 
$\hat{\bm\pi}=\hat{\bf p}+e{\bf A}({\bf r})$,
where $\hat{\bf p}$ is the electron momentum operator,
and $\mathbf{A}({\bf r})$ is the vector potential,
$m^{\ast}$ is the conduction-band mass of the electron, $e$ is the 
elementary charge, $F$ is an external electric field applied along the 
$z$-axis, $U_{conf}(\mathbf{r})$ is the confinement potential energy,
and $\hat{\mathds{1}}$ is the $2\times 2$ unit matrix.
The spin Zeeman splitting term $\hat{\mathcal{H}_Z}$ is given by
\begin{equation}
\hat{\mathcal{H}_Z} = g^{\ast}\mu_B{\bf B}\cdot\hat{\bm\sigma}
\label{HZ},
\end{equation}
where $\mu_B$ is the Bohr magneton, $g^{\ast}$ is the scalar electron effective 
Land\'e factor, $\hat{\bm\sigma}$ is the vector of the Pauli matrices. 
For the magnetic field directed along the nanowire axis, ${\bf B}=(0, 0,B)$, 
the vector potential can be chosen in the symmetric form 
${\bf A}({\bf r})=({\bf B}\times{\bf r})/2$. 
Since the nanowire which is considered within this model
is grown in the [111] direction,
the Dresselhaus spin-orbit interaction is absent for momentum along the 
nanowire
(also for the [100] nanowires, this type of the spin-orbit interaction is weak,
and $\hat{\mathcal{H}_D}$ can be neglected).\cite{Weperen_PRB91p201413y2015}
The last term in r.h.s. of Eq.~(\ref{Hamiltonian}) represents the Rashba 
interaction of the electron's spin with an electric 
field,\cite{Bandyopadhyay_book2008}
\begin{equation}
\hat{\mathcal{H}_R} =
\frac{\alpha}{\hbar} {\bf F} \cdot ( \hat{\bm\sigma} \times \hat{\bm\pi} ).
\label{HR}
\end{equation}
The Rashba parameter $\alpha$
which measures the strength of the  interaction can be given in terms of
the energy band gap $E_g$ and  the spin-orbital splitting 
$\Delta_{SO}$, as follows,\cite{Pikus1995}
\begin{equation}
\alpha =
\frac{\pi e\hbar^2}{m^{\ast}}
\frac{\Delta_{SO}(2E_g+\Delta_{SO})}{E_g(E_g+\Delta_{SO})(3E_g+2\Delta_{SO})}.
\label{Rashba-parameter}
\end{equation}

Since the electric field due to the source-drain voltage is directed along the 
axis of the nanowire, ${\bf F}=(0, 0, F)$, the Rashba Hamiltonian can be written
as
\begin{equation}
\hat{\mathcal{H}_R}
=
\frac{\alpha F}{\hbar}
\left [\begin{array}{ccc}
0 & \hat{\pi}_y+i\hat{\pi}_x\\
\hat{\pi}_y-i\hat{\pi}_x & 0\\
\end{array}\right ]
,
\label{HR1}
\end{equation}
where $\hat{\pi}_x=\hat{p}_x-eyB/2$ and $\hat{\pi}_y=\hat{p}_y-exB/2$.
For the present calculations of the magnetotransport characteristics of the 
considered nanosystems, we assume that both ends of the nanowire are attached 
through the perfect contacts to the reflectionless reservoirs of electrons 
(source and drain).
We also assume that only a small source-drain voltage is applied.
Besides the fact that within the limits of the linear response theory
the conductance in such case does not depend on the applied voltage,
it also means that
only low electric fields are present in the nanowire, and the
change of the potential profile can be neglected as well as the Rashba term.
However, we include in our calculations the effect of the intrinsic spin-orbit 
interaction which stems from the band structure by an appropriate 
renormalization of the electron Land\'e factor according to the second-order 
of the ${\bf k}\cdot{\bf p}$ perturbation theory\cite{Winkler_book2003}.

The rotational symmetry of the cylindrical nanowire allows us to split
$U_{conf}(\mathbf{r})$ into longitudinal $U_{\parallel}(z)$ and lateral 
$U_{\perp}(x,y;z)$ terms,
\begin{equation}
U_{conf}(x,y,z)
=
U_{\perp}(x,y;z)
+
U_{\parallel}(z) \; .
\label{potential}
\end{equation}
The longitudinal confinement potential energy is determined by the 
position-dependent energy of the conduction-band bottom: 
$U_{\parallel}(z)=E_c(z)$, whereas the lateral confinement potential energy is 
taken in the form of the finite potential well: $U_{\perp}(x,y;z)= U_0$ for 
$x^2(z)+y^2(z)>r^2(z)$ and $U_{\perp}(x,y;z)=0$ elsewhere, where $r(z)$ is 
the radius of the nanowire at the coordinate $z$, and $U_0$ is the height of 
the potential energy barrier.  
In the present calculations, we assume that the radius of the constriction is 
given by the formula
\begin{equation}
r(z) =
r_0 
- 
(r_0-r_c) 
\exp{ \left[ - \left(\frac{z-z_0}{L_{c}/2}\right)^2 \right] } 
\;,
\label{constriction}
\end{equation}
where $L_c=60$~nm is the length of the constriction region 
[cf. Fig.~\ref{fig:model}(a)], $z_0=100$~nm is the position of its center 
(measured with respect to the source, for which $z=0$),
the nanowire has length $L=200$~nm, $r_0=20$~nm is the 
radius of the nanowire outside the constriction region, and $r_c$ is the radius 
of the nanowire in the middle of the constriction.
Since for the values of $r_c$ less then $\sim$5~nm
the band structure strongly depends on the geometric parameters of the 
nanosystems,\cite{Niquet_PRB73p165319y2006, Carter_PRB77p115349y2008}
and thus the effective mass approximation is no longer valid,
we limit our calculations to $r_c>10$~nm.

One of the consequences of the dependence (\ref{constriction}) is the formula 
for the energy band gap in the nanowire:\cite{Yao_NL8p4557y2008}
\begin{equation}
E_g^{nano}= E_g^{bulk}+aS,
\label{Yao-formula}
\end{equation}
where $a$ is an adjustable parameter, and $S$ is the surface area to volume 
ratio (SVR). 
For the cylindrical nanowire SVR is a decreasing function of the aspect 
ratio parameter $2r(z)/L$. 
Therefore, for the geometric parameters assumed in this paper,
the energy band gap  in the region of the constriction changes its 
value by about 30\% with the change of the radius for the fixed value of the 
bulk energy gap.
This shows that the presence of the constriction in the 
cylindrical nanowire leads to the modification of the electronic properties, 
which in turn affects the spin transport.

When all the assumptions described above are taken into consideration,
the Hamiltonian takes on a simplified form,
and the Pauli equation 
can be written as 
\begin{equation}
\left [\begin{array}{ccc}
\hat{\mathcal{H}}_0+g^{\ast}\mu_BB-E&0\\
0&\hat{\mathcal{H}}_0-g^{\ast}\mu_BB-E\\
\end{array}\right ]
\left [\begin{array}{cccc}
\psi_{\uparrow}({\bf r})\\
\psi_{\downarrow}({\bf r})\\
\end{array}\right ]
=
\left [\begin{array}{cccc}
0\\
0\\
\end{array}\right ],
\label{Pauli-equation}
\end{equation}
where $E$ is the eigenenergy and $\psi_{\sigma}({\bf r})$ is the $\sigma$ 
component of the spinor $[{\sigma}=\uparrow(\downarrow)]$.
The Hamiltonian $\hat{\mathcal{H}}_0$ has the  form
\begin{equation}
\hat{\mathcal{H}}_0
=
\frac{\hat{p}^2}{2m^{\ast}}
+
\frac{1}{2}\omega_c\hat{L}_z
+
\frac{1}{8}m^{\ast}\omega_c^2(x^2+y^2)
+
U_{\perp}\big(x,y;z\big)
+
E_c(z),
\label{scalar-hamiltonian}
\end{equation}
with the cyclotron frequency $\omega_c=eB/m^{\ast}$, and the $z$-th component 
of the angular momentum operator $\hat{L}_z=x\hat{p}_y-y\hat{p}_x$. 
In our calculations, the energy is measured with respect to the 
conduction-band bottom, i.e., for the considered nanowire, which is made of
homogeneous material, we put $E_c(z)=0$.

The diagonal form of the matrix equation~(\ref{Pauli-equation}) is particularly 
useful for the calculations because it allows us to expand each of the spinor 
components in the basis of the transverse quantum states 
$\chi_{n}(x,y;z)$ for each $z$,
\begin{equation}
\psi_{\sigma}(x,y,z)
=
\sum_n
\phi_{{\sigma}n}(z)\chi_{n}(x,y;z) \;.
\label{adiabatic-basis}
\end{equation}
The coefficients $\phi_{\sigma n}(z)$ of the linear combination represent 
the longitudinal part of the component $\psi_{\sigma}(x,y,z)$ of the spinor. 
The quasi-separable form of $\psi_{\sigma}(x,y,z)$ allows us to find the 
transverse quantum states $\chi_{n}(x,y;z)$ and the corresponding transverse 
energies $E_n^{\perp}(B;z)$ by solving for each fixed $z$ the two-dimensional
Schr\"odinger equation
\begin{equation}
\hat{\mathcal{H}}_0^{\perp}\chi_n(x,y;z)
=
E_n^{\perp}(B;z)\chi_n(x,y;z)
\label{transverse-Schroedinger}
\end{equation}
with the Hamiltonian~$\hat{\mathcal{H}}_0^{\perp}$ given by
\begin{equation}
\hat{\mathcal{H}}_0^{\perp}
=
\frac{\hat{p}_x^2}{2m^{\ast}}
+
\frac{\hat{p}_y^2}{2m^{\ast}}
+
\frac{1}{2}\omega_c \hat{L}_z
+
\frac{1}{8}m^{\ast}\omega_c^2(x^2+y^2)
+
U_{\perp}\big(x,y;z\big) \;.
\label{Hamiltonian-2D}
\end{equation}
The boundary conditions are assumed in the form
$\lim_{x,y\to\infty}\chi_n(x,y;z)=0$. 
We solve Eq.~(\ref{transverse-Schroedinger}) by means of the variational 
method using the approach discussed in more detail 
in~Ref.~\onlinecite{Woloszyn_JP-CM26p325301y2014}.

Since the constant value of the spin Zeeman splitting term
is not appropriate for the description of spin-dependent transport 
phenomena in the nanosystems, 
an attempt to grasp this issue was made within the ${\bf k}\cdot{\bf p}$ 
approach.
The energy-gap dependent effective mass
is given by the relation\cite{Fabian_APS57p565y2007}
\begin{equation}
\frac{1}{m^{\ast}}
=
\frac{1}{m_0}
+
\frac{2 P^2}{3 \hbar^2}
\bigg(
\frac{2}{E_g}
+
\frac{1}{E_g+\Delta_{SO}}
\bigg),
\label{Fabian-meff-formula} 
\end{equation}
where $P=9.63$~eV$\cdot${\AA} is the parameter of the extended Kane model,
obtained for InSb from the 40-band tight-binding model 
by Jancu et al.\cite{Jancu2005}
For the parabolic approximation of the dispersion relation, the second order 
perturbation theory leads to the energy-independent expression for the effective 
Land\'e factor,\cite{Roth_PR114p90y1959, Hermann_PRB15p823y1977, 
Smith_PRB35p7729y1987, Fabian_APS57p565y2007} 
\begin{equation}
g^{\ast}
=
g
\bigg[
1+
\bigg(
1-\frac{m_0}{m^{\ast}}
\bigg)
\frac{\Delta_{SO}}{3E_g+2\Delta_{SO}}
\bigg],
\label{Roth-formula}
\end{equation}
which depends on the band gap and the spin-orbital splitting.
The symbols $g$ and $m_0$ denote the Land\'e factor and the rest mass of the 
free electron in vacuum, respectively.
In the present paper, the parabolic dispersion relation is assumed
since we consider transport only through the lowest transverse state
($E_{0}^{\perp} < 150$~mV).
This means that even for the relatively strong magnetic field used in our
calculations (up to 8~T) we have
$|g^{\ast} \mu_B B|<25$~meV, and the energies of the electrons are within the
range which can be well approximated by the parabolic 
relation.\cite{Chelikowsky1984,Kim2009}

For the nanostructures, Eq.~(\ref{Roth-formula}) requires some modification 
to be consistent with formula~(\ref{Yao-formula}).  
This problem has been addressed in Ref.~\onlinecite{Kiselev_PRB58p16353y1998}, 
where the authors presented the procedure of including the quantum size effect.  
In line with their work, we modify the formula for the effective Land\'e 
factor of the nanowire as follows:
\begin{equation}
g^{\ast}(B; z)
=
g
\bigg[
1+
\bigg(
1-\frac{m_0}{m^{\ast}}
\bigg)
\frac{\Delta_{SO}}{3[E_g+E_{0}^{\perp}(B;z)]+2\Delta_{SO}}
\bigg].
\label{ws-formula1}
\end{equation}
The term $E_g+E_{0}^{\perp}(B;z)$,
with the lowest transverse-state energy level $E_{0}^{\perp}(B;z)$,
can be understood as the 
magnetic-field and position-dependent band 
gap in the nanowire, $E_g^{nano}$, because it depends on both the geometric 
parameters of the considered nanosystem and the magnetic field.
 \begin{figure}
 \begin{center}
 \includegraphics[width=0.50\textwidth]{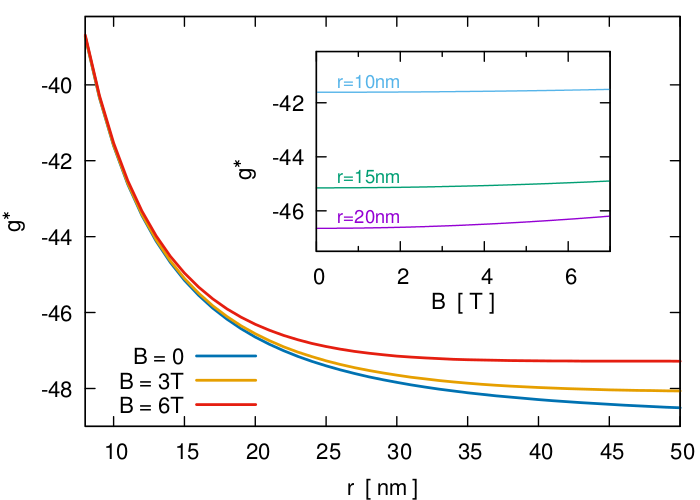}
 \caption{
 Effective Land\'e factor $g^{\ast}$ as a function of the nanowire radius $r$.
 in the presence of the magnetic field $B=0$, $3$~T, and $6$~T.
 Insets show $g^{\ast}$ as a function of the magnetic 
 field $B$ calculated for the nanowire with radius $r$ equal to
 10~nm, 15~nm, and 20~nm.}
 \label{fig:Lande}
 \end{center}
 \end{figure}
The effective Land\'e factor calculated from formula~(\ref{ws-formula1})
is presented in Fig.~\ref{fig:Lande} for the following parameters of InSb: 
$E_g=0.235$~eV, $\Delta_{SO}=0.81$~eV.
The results exhibit interesting features that seem to be important for 
the $g^{\ast}$-factor engineering, namely, in all the cases the effective 
Land\'e factor
decreases with the increasing  nanowire radius,
and the limiting values of $g^{\ast}$ obtained for large radius $r$ 
approach the bulk values of the Land\'e factor for InSb. 
The dependence of the electron Land\'e factor on the magnetic field 
is shown in the inset of Fig.~\ref{fig:Lande}
for the center of the constriction (where the nanowire radius is $r=r_c=10$~nm),
and in the right inset of Fig.~\ref{fig:Lande} for the region outside the 
constriction (i.e., for $r=r_0=20$~nm). 
The magnetic-field effect on the effective 
Land\'e factor is more pronounced outside the constriction, and in general for 
the nanowires with larger diameters.
For a given radius, the electron Land\'e factor which corresponds to the 
lowest-energy transverse mode
only weakly depends on the magnetic field $B$.  
This means that its value is determined mainly by the geometric parameters of 
the constriction, whereas the magnetic field can be regarded as a weak 
perturbation.
The diameter-dependence of the electron Land\'e factor affects the spin Zeeman 
splitting, making it a non-linear function of the magnetic field due to the 
presence of the constriction. 
Those properties are consistent with
the basic properties of the electron Land\'e factor which were determined within 
a more advanced model based on the Ogg-McCombe effective Hamiltonian that 
includes the non-parabolicity and anisotropy 
effects.\cite{Lopez_IoP-CM20p175204y2008}

The longitudinal part $\phi_{\sigma n}(z)$ of the spinor component 
$\psi_{\sigma}(x,y,z)$ satisfies the inhomogeneous differential equation in the 
form\cite{Woloszyn_JP-CM26p325301y2014}
\begin{equation}
\bigg[
-\frac{\hbar^2}{2m^{\ast}}\frac{d^2}{dz^2}
+ E_n^{\perp}(B;z)
- E 
\pm g^{\ast}_n(B; z)\mu_BB
\bigg]\phi_{\sigma n}(z)
=
\sum_{n^{\prime}}
\Lambda_{nn^{\prime}}(z)\phi_{\sigma n^{\prime}}(z) \;.
\label{longitudinal-Pauli}
\end{equation}
The matrix elements $\Lambda_{nn^{\prime}}(z)$ represent the coupling between 
the transverse modes.
Since
in the present calculations we assume that only the lowest-energy transverse 
state is occupied, we can put the right-hand side of 
Eq.~(\ref{longitudinal-Pauli}) equal to zero,
and solve it only for $n=0$ to find the transmission coefficient
$T_{\sigma}(E;B)$.

The magnetotransport properties of the considered nanosystem can be 
quantitatively characterized by the magnetoresistance (MR), which is defined as 
the ratio of the resistance change due to the magnetic field, $R(B)-R(0)$, to 
the resistance measured in zero magnetic field, $R(0)$, i.e.,
\begin{equation}
\mathrm{MR}(B)=\frac{R(B)-R(0)}{R(0)}\;.
\label{magnetoresistance}
\end{equation}
The resistance $R$ of the nanowire is calculated as the inverse of its 
conductance $G$, which is given by the sum of two spin-dependent 
contributions, $G=G_{\uparrow}+G_{\downarrow}$. 
This is a direct consequence of the Mott two-current 
model\cite{Mott_PRSA153p699y1936} which is applied here.
In turn, the conductance is calculated as the ratio of the spin-dependent 
electric current $I_{\sigma}(B)$ to the voltage $V$ applied between the
ends of the nanowire, i.e., $G_{\sigma}(B)=I_{\sigma}(B)/V$.
In general, the spin-dependent current can be calculated from
the formula\cite{DiVentra_book2008, Ferry_book2009}
\begin{equation}
I_{\sigma}(B) = 
\frac{e}{h} \int_{0}^{\infty}dE\; 
T_{\sigma}(E;B)[f_S(E;\mu_S)-f_D(E;\mu_D)],  
\label{Landauer-formula}
\end{equation}
where $T_{\sigma}(E;B)$ 
is the spin-dependent transmission coefficient, and 
$f_{S(D)}(E;\mu_{S(D)})$ is the Fermi-Dirac distribution function for the 
electrons in the source~$S$ (drain~$D$) with the electrochemical potential 
$\mu_{S(D)}$.
The electrochemical potentials are given by $\mu_S=E_F$ and $\mu_D=E_F-eV$, 
where the Fermi energy $E_F$ is assumed to be the same for the source and 
the drain.
In the case of low temperature and low voltage,
the conductance takes on the simple form:
\begin{equation}
G_{\sigma}(B)
=
\frac{e^2}{h} T_{\sigma}(E_F;B).
\label{conductance}
\end{equation}
Finally, the total  resistance of the nanowire in the presence of the magnetic 
field within the two-current model is given by
\begin{equation}
R_{\textrm{total}}(B)
=
\frac{R_{\uparrow}(B)R_{\downarrow}(B)}
{R_{\uparrow}(B)+R_{\downarrow}(B)},
\label{resistance2model}
\end{equation}
where $R_{\uparrow (\downarrow)}=1/G_{\uparrow (\downarrow)}$.
Using Eqs.~(\ref{magnetoresistance}) and (\ref{resistance2model}), we determine 
the MR of the nanowire.

\section{\label{sec:results}Results and discussion}

We have applied the methods presented in the previous section to determine the 
spin-dependent magnetotransport in the InSb nanowire with the 
geometric constriction
in the limit of the low electric field and the low temperature.
This means that we assume the coherence of the electronic transport within the 
linear response theory, which is sufficient for the theoretical description of 
the majority of transport experiments in the semiconductor nanowires in terms 
of the conductance. 

The constriction in the nanowire creates an effective potential 
barrier.\cite{Woloszyn_JP-CM26p325301y2014}
The additional effect due to the magnetic field is associated with the 
elimination of the spin degeneracy of the longitudinal electron states due to 
the spin Zeeman effect.  
It implies that the effective shape of the potential barrier 
created by the constriction depends on the electron spin state.
In the InSb nanowire, the potential barrier for the electrons with 
spin up ($\uparrow$) is lower than for the electrons with spin down 
($\downarrow$).
Therefore, the transmission coefficients $T_{\uparrow}$ and $T_{\downarrow}$ 
are different, and the corresponding spin-dependent conductances also 
differ, which can be easily demonstrated using 
the concept of the spin conductance defined as:
\begin{equation}
\Delta G(B)
=
G_{\uparrow}(B)-G_{\downarrow}(B)\;.
\label{spin-conductance}
\end{equation}

Let us first investigate the influence of the constriction radius on the spin 
conductance in the presence of the magnetic field for the fixed radius of the 
nanowire ($r_0=20$~nm) and for the Fermi energy $E_F=50$~meV. 
In this case, the coherent propagation of electrons is limited to only one 
transport channel, labeled by $n=0$ in Fig.~\ref{fig:model}(b). 
\begin{figure}
\begin{center}
\includegraphics[width=0.50\textwidth]{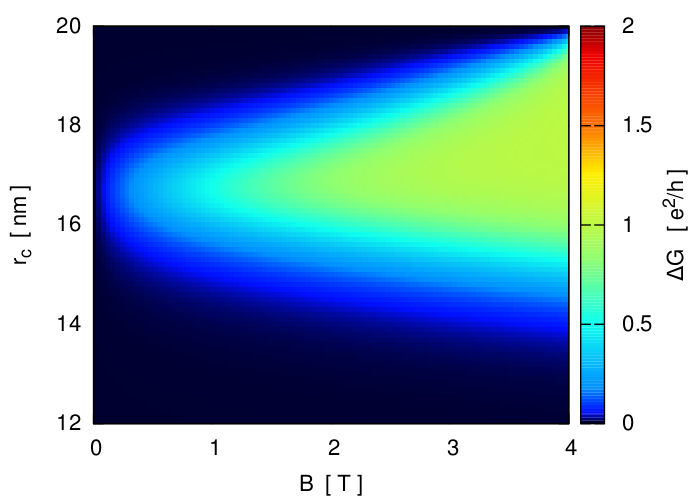}
\caption{Spin conductance $\Delta G$ as a function of the constriction radius 
$r_c$ and the magnetic field $B$ for the Fermi energy $E_F=50$~meV.}
\label{fig:spin-conductance}
\end{center}
\end{figure}
The results presented in Fig.~\ref{fig:spin-conductance} indicate that
the spin conductance $\Delta G$ is nearly zero in the two ranges of 
the constriction radius, in which the spin-dependent conductances 
in both the spin channels are almost the same:
(i)~in the region of the small constriction radius ($r_c<14$~nm)
$G_{\uparrow} \cong G_{\downarrow} \cong 0$, (ii)~for $r_c$ near $20$~nm
$G_{\uparrow} \cong G_{\downarrow} \cong 1$. 
The non-zero values of the spin conductance  correspond to 
$G_{\uparrow}\neq G_{\downarrow}$.
In this case $G_{\uparrow}$ is nearly constant, while
$G_{\downarrow}$ decreases as the magnetic field increases. 
This is a consequence of the increasing role of the spin Zeeman effect for the 
higher magnetic fields.

The above observation allows us to 
propose a possible application of the InSb nanowire with the constriction in
spintronics.
It can operate as a spin 
filter\cite{Wojcik_PRB86p165318y2012, Wojcik_APL2012p242411y2013} in which the 
spin filtering operation results from the joint effect of the constriction and
spin Zeeman effect controlled by the magnetic field.
This operation is similar to the operation of the quantum point 
contacts.\cite{Bezak_AP322p2603y2007,Wojcik_JAP118p014302y2015}

\begin{figure}
\begin{center}
\includegraphics[width=0.50\textwidth]{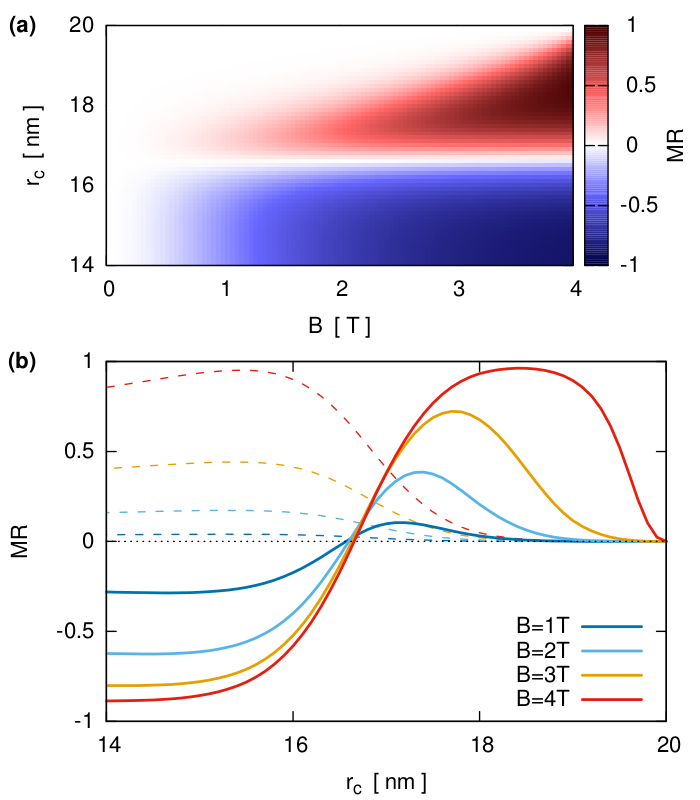}
\caption{Magnetoresistance as a function of:
(a)~radius $r_c$ of the constriction and magnetic field $B$,
(b)~radius $r_c$ of the constriction  
(cross sections of (a) at different $B$;
dashed lines correspond to the results obtained when spin of the electrons 
is neglected).}
\label{fig:mgr}
\end{center}
\end{figure}
The results of the calculations of MR based on Eqs.~(\ref{magnetoresistance}) and 
(\ref{resistance2model}) are presented in Fig.~\ref{fig:mgr}.
They show that the sign of MR depends on the radius of 
the constriction, $r_c$, and that -- for each $B$ -- the change of sign 
of MR takes place in a very narrow range, around $r_c\approx 16.8$~nm.
If the radius of the constriction decreases 
starting from $r_c=20$~nm, which corresponds to a nanowire without constriction,
the MR gradually increases, and its sign is positive
[cf.Fig.~\ref{fig:mgr}(b)].
The rate of change of the MR as a function of the applied
magnetic field, and its maximum value together with the position of this 
maximum depend on the magnetic field.
If the magnetic field reaches the value of 4~T, the MR is close 
to one, with maximum at $r_c \approx 18.5$~nm.
At lower magnetic fields the maxima of MR are lower, and their position is 
shifted towards the smaller values of $r_c$.
A further reduction of the constriction radius leads to the fairly rapid 
decrease of MR, which becomes negative for $r_c \lesssim 16.8$~nm. 
For $r_c \lesssim 15$~nm, the negative MR saturates at some 
constant level, which is 
magnetic-field dependent, e.g., the minimum of about $-0.9$ is reached for 
$B=4$~T.
Noteworthy, the MR calculated for spinless electrons is always 
non-negative [cf. dashed lines in Fig.~\ref{fig:mgr}(b)]. 
On the other hand, using suitably chosen constant value of the effective Land\'e
factor allowed us to reproduce the change of the MR sign
in the case of the geometric and material parameters used in the present 
calculations, and for the transport only via the lowest transverse state,
but at the cost of underestimated positive MR values obtained from such 
simplified approach.

\begin{figure}
\begin{center}
\includegraphics[width=0.50\textwidth]{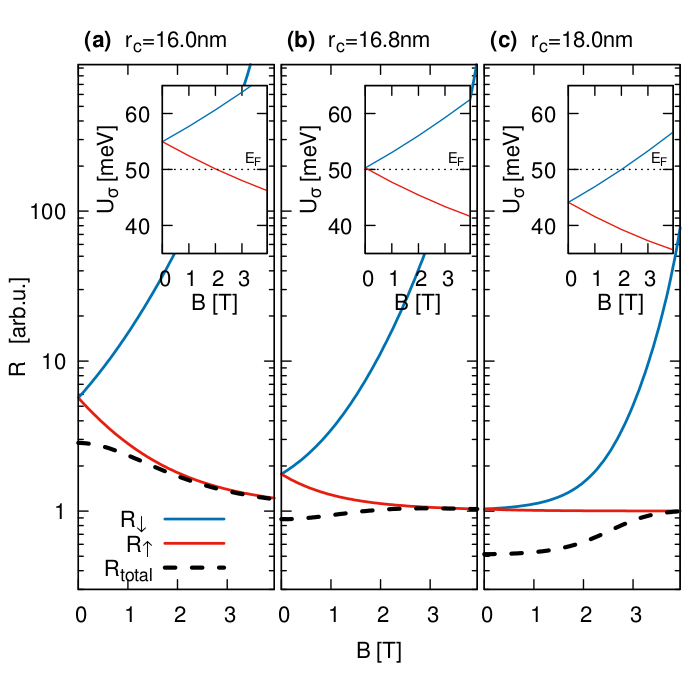}
\caption{Spin-up ($R_{\uparrow}$, red curves), spin-down ($R_{\downarrow}$, blue
curves) and total 
($R_{\textrm{total}}$, dashed lines) resistances of the nanowires for three 
different constriction radii $r_c$: (a)~16~nm, (b)~16.8~nm, (c)~18~nm.
Insets show the effective heights of barriers at the centers of constrictions,
i.e., $U_{\sigma}=E_{0}^{\perp}(B;z_0) \pm g^{*}(B;z_0) \mu_B B$,
and $E_F$ is the Fermi energy.}
\label{fig:Rfig}
\end{center}
\end{figure}

The physical interpretation of the positive/negative magnetoresistance 
transition (Fig.~\ref{fig:mgr}) can be given based on the results presented in 
Fig.~\ref{fig:Rfig}.
As mentioned above,
the constriction in the nanowire creates the effective potential barrier
for the conduction electrons, which strongly affects the transport
of the electrons through the nanowire by changing the transmission coefficient.
The height of this barrier at the center of the constriction, i.e., at $z=z_0$, 
is given by $U_{\sigma}(B) = E_0^{\perp} (B; z_0) \pm g^{*}(B; z_0) \mu_0B$ 
(insets in Fig.~\ref{fig:Rfig}).

For $B=0$ the barrier height and the transmission coefficient
are independent of the electron spin state; therefore,
the only effect of the decreasing constriction radius
is the increase of the potential barrier height, which causes
that the transmission coefficient is reduced and the resistance increases.
For $B>0$ the  spin degeneracy is lifted and the potential barrier height
$U_{\sigma}(B)$  increases (decreases) with increasing $B$ for spin down 
(spin up) electrons (cf. insets of Fig.~\ref{fig:Rfig}).

Let us consider the effect of the narrowing of the constriction on the 
magnetoresistance.
For the constriction radii from the interval $17 \; \mathrm{nm} < r_c < 20$~nm 
and for either spin, $U_{\sigma}(B=0) < E_F$. 
The spin-up barrier height $U_{\uparrow}(B)$  decreases with increasing $B$ 
[Fig.~\ref{fig:Rfig}(c)], which causes that the transmission probability for the
spin-up electrons is close to 1 and becomes independent of $B$.
As a result, resistance $R_{\uparrow}$ is constant [cf. Fig.~\ref{fig:Rfig}(c)].
On the other hand, potential barrier height $U_{\downarrow}$ increases with $B$,
which reduces the transmission probability for the spin-down electrons and
enlarges $R_{\downarrow}$.
As a consequence, the total resistance increases with the increasing magnetic 
field [cf. Fig.~\ref{fig:Rfig}(c)], which leads to the positive 
magnetoresistance for $r_c > 17$~nm.

For $r_c = 16.8$~nm [Fig.~\ref{fig:Rfig}(b)], $U_{\sigma}(B=0) = E_F$.
If the magnetic field increases, $U_{\uparrow}(B)$ and $U_{\downarrow}(B)$ 
change at approximately the same rates (but with the opposite slopes).  
Therefore, the growth of $R_{\downarrow}$ is compensated by the drop of 
$R_{\uparrow}$.
As a result, the total resistance is nearly constant as a function of the 
magnetic field and the magnetoresistance tends to zero.
The positive/negative magnetoresistance transition occurs in the very narrow 
interval of the constriction radii around $r_c = 16.8$~nm and is almost 
independent of the magnetic field [cf. Fig. ~\ref{fig:mgr}(b)].  
This feature results from the weak dependence of the total resistance on the 
magnetic field [Fig.~\ref{fig:Rfig}(b)].

For $r_c < 16.8$~nm, $U_{\sigma}(B=0) > E_F$  and $U_{\sigma}(B)$ increases 
(decreases) with increasing $B$ for spin-down (spin-up) electrons 
[Fig.~\ref{fig:Rfig}(a)], which leads to the increase of $R_{\downarrow}(B)$ and 
decrease of $R_{\uparrow}(B)$.
Since the $R_{\downarrow}(B)$ is large for $B > 0$, according to Eq. 
(\ref{resistance2model}), it does not affect  the total resistance considerably.
Then, the change of the total resistance is determined mainly by $R_{\uparrow}$ 
and decreases with increasing $B$.
Therefore, the magnetoresitance is negative for the sufficiently narrow 
constriction [Fig.~\ref{fig:mgr}(b)].

\section{\label{sec:concl}Concluding remarks}
We have studied the spin-dependent magnetotransport of the 
semiconductor cylindrical nanowires with the geometric constriction in the 
presence of the magnetic field applied parallel to the nanowire axis. 
The results have been obtained within the three-dimensional model of the 
nanowire using the adiabatic approximation, with the transverse states 
calculated by the variational method.
The associated $z$-dependent transverse-state energies create the 
effective-potential barriers and modify the Land\'e factor,
making it the position- and the magnetic-field dependent quantity.
The effective $g$-factor is a monotonically decreasing function of the nanowire
radius,
and
the effect of the magnetic field on the effective $g$-factor is negligibly small 
in the constriction region for the lowest-energy transverse mode.
Using the two-current Mott model, we have investigated the influence of the 
constriction radius and the magnetic field on the spin conductance in the 
coherent regime of the transport.
We have shown that the sign of magnetoresistance can be reverted by changing
the radius of the constriction, which strongly affects the transverse states.
On the contrary, the increase of the
magnetic field while the radius of the constriction is kept constant
leads to the increase of the magnetoresistance but does not 
change its sign. 
We have explained the positive/negative magnetoresistance transition as a 
combined result of the squeezing of the tranverse electron states in the region 
of the constriction and the spin Zeeman splitting. 

Finally, we want to point out that the geometric inhomogeneity, which is 
represented in our calculations by the single constriction, can be used
as a model of either an intentionally fabricated change of the nanowire radius, 
or a ring-shaped gate.
The present results indicate that the InSb nanowire with the constriction
can operate as a spintronic nanodevice in the coherent regime of the electronic 
transport, e.g., the intentionally introduced constriction can serve as a 
tunnel junction enhancing the spin polarization of the current flowing through
the nanowire.\cite{Rashba_PRB62pR16267y2000} 
The anomalous properties of the magnetoresistance, demonstrated in the present 
paper, can be applied in spintronics, e.g., for modifying the resistance of the
spin current, or in sensor technology, e.g., for detecting the inhomogeneity
of the nanowire and estimating its size.

\begin{acknowledgments}
This project is supported by the National Science Centre, Poland under 
grant DEC-2011/03/B/ST3/00240.
\end{acknowledgments}


%

\end{document}